\documentclass[twocolumn,showpacs,preprintnumbers,amsmath,amssymb]{revtex4}
\usepackage{graphicx}% Include figure files
\usepackage{dcolumn}% Align table columns on decimal point
\usepackage{bm}% bold math
\usepackage{color}
\begin{document}

\title{Vector Rogue Waves and Baseband Modulation Instability in the Defocusing Regime}

\author{Fabio Baronio$^1$, Matteo Conforti$^2$, Antonio Degasperis$^3$, Sara Lombardo$^4$, Miguel Onorato$^5$, and Stefan Wabnitz$^1$}
\affiliation{$^1$ Dipartimento di Ingegneria dell'Informazione, Universit\`a di Brescia, Via Branze 38, 25123 Brescia, Italy\\
$^2$ PhLAM/IRCICA UMR 8523/USR 3380,  CNRS-Universit\'e Lille 1,  F-59655 Villeneuve d'Ascq, France
\\
 $^3$ INFN, Dipartimento di Fisica, ``Sapienza'' Universit\`a di Roma, P.le A. Moro 2, 00185 Roma, Italy
 \\ $^4$ Department of Mathematics and Information Sciences, Northumbria University, Newcastle upon Tyne NE2 1XE, United Kingdom
 \\ $^5$ Dipartimento di Fisica, Universit\`a di Torino, via P. Giuria, 10125 Torino, Italy}

\date{\today}% It is always \today, today,

\begin{abstract}
%We show and analyze exact explicit rogue wave solutions of the vector nonlinear Schr\"odinger equation, in defocusing regimes.
We report and discuss analytical solutions of the vector nonlinear Schr\"odinger equation that describe rogue waves in the defocusing regime.
This family of solutions includes bright-dark and dark-dark rogue waves. 
%We evince that rogue waves'  existence is bound to induced MI conditions, in particular to baseband MI.
The link between modulational instability (MI) and rogue waves is displayed
by showing that only a peculiar kind of MI, namely baseband MI, can sustain rogue wave formation.
The existence of vector rogue waves in the defocusing regime is expected to be a crucial progress in explaining extreme
 waves in a variety of physical scenarios described by multi-component systems, from oceanography to optics and plasma physics.  
\end{abstract}

\maketitle

\textit{Introduction.--}
Rogue waves are extremely violent phenomena in the ocean: an encounter with such a wave can be disastrous even to big ocean liners. These waves can be also very dangerous for various hydrotechnic constructions. 
This makes the study of  rogue waves a very important problem. Hence, it is not surprisings that the phenomenon 
of rogue waves has attracted ample attention of oceanographers in the last decade \cite{hopkin04,muller05,perkin06,kharif09}.
However, although the existence of rogue waves has now been confirmed by multiple observations, 
uncertainty remains on their fundamental origins. This hampers systematic approaches to study their characteristics,
including the predictability of their appearance \cite{pelinosky2008}.

The research on rogue waves in oceans has attracted recently the attention of researchers in many other fields in physics \cite{akhmediev2010sp}.
Rogue waves have been observed in nonlinear optics and lasers \cite{erkin2009}, atmosphere \cite{stenflo2009}, 
plasma physics \cite{bailung11}, and matter waves (Bose-Einstein condensate) \cite{bludov2009}.

The possibility to reach a general understanding of rogue-wave formation is still an open question \cite{akhmediev2010sp}. 
Nonetheless, the ongoing debate stimulates the comparison of predictions and observations between distinct topical areas, 
in particular hydrodynamics and nonlinear optics \cite{onorato13}, in situations where analogous dynamical behaviors can be identified through the use of common mathematical models. 

So far, the focusing nonlinear Schr\"odinger equation (NLSE) has played a pivotal role as a universal model for rogue wave solutions. The Peregrine soliton, predicted $30$ 
years ago \cite{peregrine83}, is the simplest rogue-wave solution associated with the focusing NLSE, and it has been recently experimentally observed in optical fibers \cite{kibler10}, water-wave tanks \cite{amin11}, and plasmas \cite{bailung11}. 
The Peregrine soliton turns out to be just the first of an infinite hierarchy of higher-order localized soliton solutions
of the focusing NLSE \cite{ACA10,guo12,amin12}.

%To note that the Ma solitons and Ackhmediev Breathers,
%whose limits lead to Peregrine solitons, are illustrative of the iduced modulation instability.
%

While rogue wave investigations are flourishing in several fields of science, moving beyond the standard focusing
NLSE description in order to model more general and important classes of physical systems is both relevant and necessary.
In
this direction, recent developments consist in i) including
dissipative terms, since a substantial supply of energy
(f.i., wind in oceanography) is generally required
to drive rogue wave formation \cite{lecap12}, or in ii) including
higher-order perturbation terms such as in the Hirota equation and in
the Sasa-Satsuma equation \cite{ankiew10,bandel12}, because of the high
amplitude or great steepness of a rogue wave, or in iii)
considering wave propagation in 2+1 dimensions as for
the Davey-Stewartson equation \cite{ohta12}.
Additional important
progress has been recently obtained by extending the
search for rogue wave solutions to coupled-wave systems,
since numerous physical phenomena require modeling waves
with two or more components in order to account for different
modes, frequencies, or polarizations. When compared to
scalar dynamical systems, vector systems may allow for
energy transfer between their additional degrees of freedom,
which potentially yields rich and significant new families
of vector rogue-wave solutions. Indeed, rogue-wave families
have been recently found as solutions of the focusing
vector NLSE (VNLSE)\cite{baronio12,liu2013,zhai2013}, 
the Three Wave Resonant Interaction equations \cite{baronio13,dega13}, 
the coupled Hirota equations \cite{chen13}, and the Long-Wave-Short-Wave resonance \cite{chen14}. 

It is a well established fact that, for the scalar NLSE,
the focusing nonlinear regime is a prerequisite for the
emergence of regular or random rogue waves (see, f.i., discussion in \cite{akhmediev2010sp}). 
To the contrary, in the scalar case the defocusing
nonlinear regime does not allow for rogue wave
solutions, even of dark nature. In coupled-wave systems,
is the focusing regime still a prerequisite for the existence of
rogue wave solutions? Or it possible to find examples of
rogue waves in defocusing regimes?

%In fact, some questions, on the relevance of focusing (defocusing) regimes as prerequisite for the emergence of rogue waves, are still opened in the analysis of nonlinear coupled systems. In particular, 
%is it possible the existence of rogue wave solutions in defocusing regimes? 

Additionally, what are the conditions under which modulation instability (MI) may produce an extreme wave event?  Indeed, it is generally recognized that MI is among the several mechanisms
which generate rogue waves \cite{peregrine83}.
A rogue wave may be the result of MI, but not every kind of MI leads to rogue
wave generation \cite{ruderman10,sluniaev10,kharif10}. 
%Hence, the open question about the conditions under which this instability may spawn an extreme wave event is important \cite{kharif10}.

In this Letter, we prove the existence of rogue wave solutions of the VNLSE in the \textit{defocusing} regime. 
%We evince that the existence of these rogue waves, in defocusing regimes, is stricktly bound to the induced MI, but not every MI conditions may spawn rogue waves.  
We evince that MI is a necessary but not sufficient condition for the existence of rogue waves. In fact, rogue waves can exist if and only if the MI gain band also contains the zero-frequency perturbation as a limiting case (baseband MI).

%Let us recall that the scalar NLS equation does not admit single dark-rogue-wave solutions,
%even in the case of a defocusing nonlinearity. Thus a major motivation in the analysis of nonlinear coupled systems is to
%find newrational solutions with a dark-rogue-wave counterpart that are relevant to practical physical systems.

%
%

\textit{Defocusing VNLSE and Rogue Waves.--}\label{sec2}
%
%Waves are assumed to be modeled by the dimensionless VLNSE:
We consider the VNLSE (also known as Manakov system) which we write in the following dimensionless form:

\begin{equation}\label{VNLS}
\begin{array} {lll}
iE^{(1)}_t+ E^{(1)}_{xx} -2 s ( |E^{(1)}|^2 +  |E^{(2)}|^2 ) E^{(1)} & = & 0, \\
iE^{(2)}_t+ E^{(2)}_{xx} -2 s ( |E^{(1)}|^2  + |E^{(2)}|^2 ) E^{(2)} & = & 0, 
\end{array} 
\end{equation}
where $E^{(1)}(x,t),\,E^{(2)}(x,t)$ represent the wave envelopes and $x,t$ are the transverse and longitudinal 
coordinates, respectively. Each subscripted variable in Eqs. (\ref{VNLS}) stands for partial differentiation. 
It should be pointed out that the meaning of the dependent variables $E^{(1)}(x,t),E^{(2)}(x,t)$, and of the coordinates $x,t$ depends on the particular applicative context (f.i., nonlinear optics, water waves, plasma physics). 

%$s$ is a real parameter. We consider $s=\pm 1$ (i.e., integrable Manakov system). 
We have normalized Eqs. (\ref{VNLS}) in a way such that $s=\pm 1$.
Note that in the case $s=-1$, Eqs. (\ref{VNLS}) refer to the focusing 
(or anomalous dispersion) regime; in the case $s=1$, Eqs. (\ref{VNLS}) 
refer to the defocusing (or normal dispersion) regime.

Like the scalar NLSE, also the focusing VNLSE (\ref{VNLS})possesses rogue wave solitons  \cite{baronio12,liu2013,zhai2013}. 
%Nonetheless, defocusing VNLSE possesses rational solutions with the property of representing amplitude 
Unlike the scalar case, and far from being obvious, we find that rational solutions
of the defocusing VNLSE do indeed exist, with the
property of representing amplitude peaks which are localized
in both $x$ and $t$ coordinates. These solutions are constructed
by means of the standard Darboux dressing method \cite{spiegazione,deg09} and, for
Eqs. (\ref{VNLS}) with $s=1$, they can be expressed as:

%\begin{subequations}\label{pere}
\begin{equation}\label{pere}
E^{(j)}=
  E_0^{(j)} \big[\frac{p^2 x^2+p^4 t^2+p x(\alpha_j+\beta \theta_j)-i \alpha_j p^2 t+ \beta \theta_j}{p^2x^2+p^4t^2+ \beta (px+1)}  \big] 
\end{equation}
%\begin{equation} \label{E2}
%E^{(2)}=  \big[a_2+\frac{2 i (k^*-k) v^* v_2}{|v|^2-|v_1|^2-|v_2|^2}  \big] e^{-i(q x+ \nu t)} 
%%
%\end{equation}
%
%\end{subequations}
%%% not similartly to the focusing, qui c'è condizione di esistenza
%
%
where
%
%%a_1 e^{i(q_1 x- \nu_1 t)} 
%
%\begin{gather*}\label{def}
\begin{equation}\label{fondo}
E_0^{(j)} =a_j e^{i(q_j x- \nu_j t)}, \; 
\nu_j=q_j^2+ 2 (a_1^2+a_2^2),\; j=1,2; \ \ \ \ \ \ \ \ \ \ \ \ \ \ \ \ \ \ \ \ \ \ \ \ \ \ \ \ \ \  \\
\end{equation}
%\end{gather*}
represent the backgrounds of expressions (\ref{pere}),
\begin{gather*}\label{def}
\alpha_j=4p^2/(p^2+4q_j^2),
\theta_j=(2q_j+ip)/(2q_j-ip), j=1,2;  \ \ \ \ \ \ \ \ \  \ \ \ \ \ \ \ \ \ \ \ \ \ \ \ \ \ \ \ \ \ \ \ \ \ \ \ \ \ \ \ \ \ \ \ \ \ \ \ \ \ \ \ \ \ \ \ \ \ \ \ \\ 
\beta=p^3/\chi(p^2+4q_1q_2),  
p=2 \textrm{{Im}}(\lambda+k), \ \ \ \ \ \ \ \ \  \ \ \ \ \ \ \ \ \ \ \ \ \ \ \ \ \ \ \ \ \ \ \ \ \ \ \ \ \ \ \ \ \ \ \ \ \ \ \ \ \ \ \ \ \ \ \ \ \ \ \  \\
q_1+q_2=2 \textrm{{Re}}(\lambda+k),  
q_1-q_2=2q, \chi=\textrm{{Im}} k.  \ \ \ \ \ \ \ \ \  \ \ \ \ \ \ \ \ \ \ \ \ \ \ \ \ \ \ \ \ \ \ \ \ \ \ \ \ \ \ \ \ \ \ \ \ \ \ \ \ \ \ \ \ \ \ \ \ \ \ \  
%
%
%
%
%
%z_n=z+V_n t
\end{gather*}
As for the computation of the complex value of $k$ and $\lambda$,
$k$ is either one of the complex solutions of the fourth order polynomial: 
\begin{equation} \label{eqk}
k^4+D_3 k^3+ D_2 k^2+ D_1 k +D_0=0,
\end{equation}
with
\begin{gather*}\label{def}
D_0 = (q^2 - a_1^2 - a_2^2)^3/(2^4 q^2) - (3/4)^3 (a_2^2 - a_1^2)^2;  \ \ \ \ \ \ \ \ \ \ \ \ \  \\
D_1 = -9(a_2^2 - a_1^2)(2 q^2 + a_1^2 + a_2^2)/(2^4 q);  \ \ \ \ \ \ \ \ \ \ \ \ \  \ \ \ \ \ \ \ \ \ \ \ \ \  \\
D_2 = -[8q^4 - (a_1^2 +a_2^2)^2 +20 q^2(a_1^2 +a_2^2)]/(2^4 q^2);  \ \ \ \ \ \ \ \ \ \ \ \ \  \\
D_3 = (a_2^2 - a_1^2)/(2q);  \ \ \ \ \ \ \ \ \ \ \ \ \  \ \ \ \ \ \ \ \ \ \ \ \ \  \ \ \ \ \ \  \ \ \ \ \ \   \ \ \ \ \ \ \ \ \ \ \ \ \  \ \ \ \ \ \ \ \ \ \ \ \ \ 
\end{gather*}
and $\lambda$ is the double solution of the polynomial:
\begin{equation} \label{eql}
\lambda^3+ A_2 \lambda^2+ A_1 \lambda +A_0=0, 
\end{equation}
with
\begin{gather*}\label{def} 
A_0=-k^3 + k(q^2 + a_1^2 + a_2^2) + q (a_2^2 -a_1^2); \ \ \ \ \ \ \ \ \ \ \ \ \  \ \ \ \ \ \ \  \ \ \ \ \ \ \ \ \ \ \ \ \  \\
A_1 = -k^2 - q^2 + a_1^2 + a_2^2; \ \ \ \ \ \ \ \ \ \ \ \ \  \ \ \ \ \ \ \ \ \ \ \ \ \   \ \ \ \ \ \ \ \ \ \ \ \ \  \ \ \ \ \ \ \ \ \ \ \ \ \  \\
A_2 = k. \ \ \ \ \ \ \ \ \ \ \ \ \  \ \ \ \ \ \ \ \ \ \ \ \ \   \ \ \ \ \ \ \ \ \ \ \ \ \  \ \ \ \ \ \ \ \ \ \ \ \ \  \ \ \ \ \ \  \ \ \ \ \ \ 
%%11
%%11
%
\end{gather*}
The expressions reported above depend on the real parameters $a_1, a_2$ and $q$ {which} originate from the naked {solution (\ref{fondo})}, namely from the backgrounds:
$a_1, a_2$ represent the amplitudes, and $2q$ the ``frequency'' difference of the waves. 

\begin{figure}[h]
\begin{center}
\includegraphics[width=7cm]{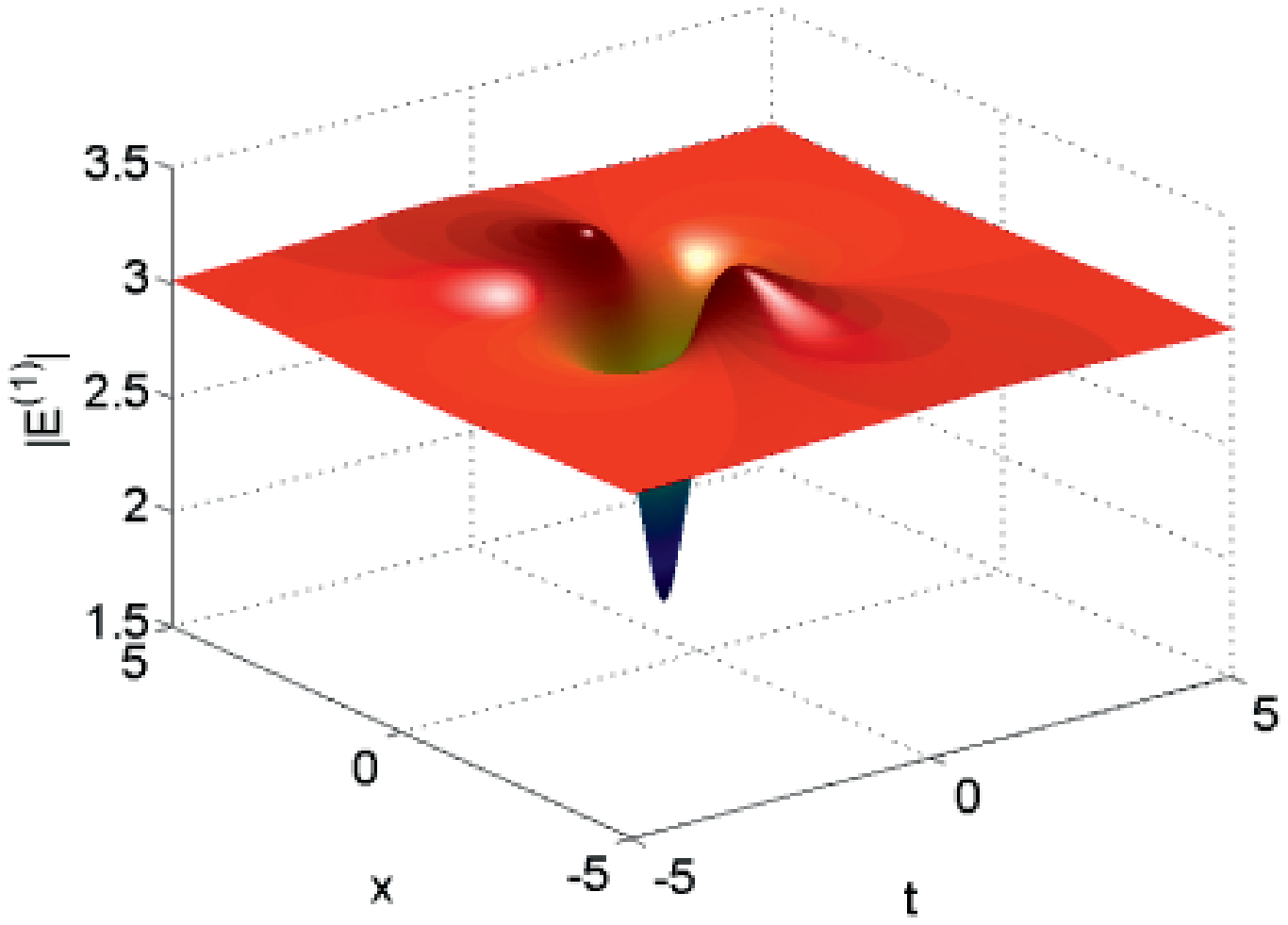}
\includegraphics[width=7cm]{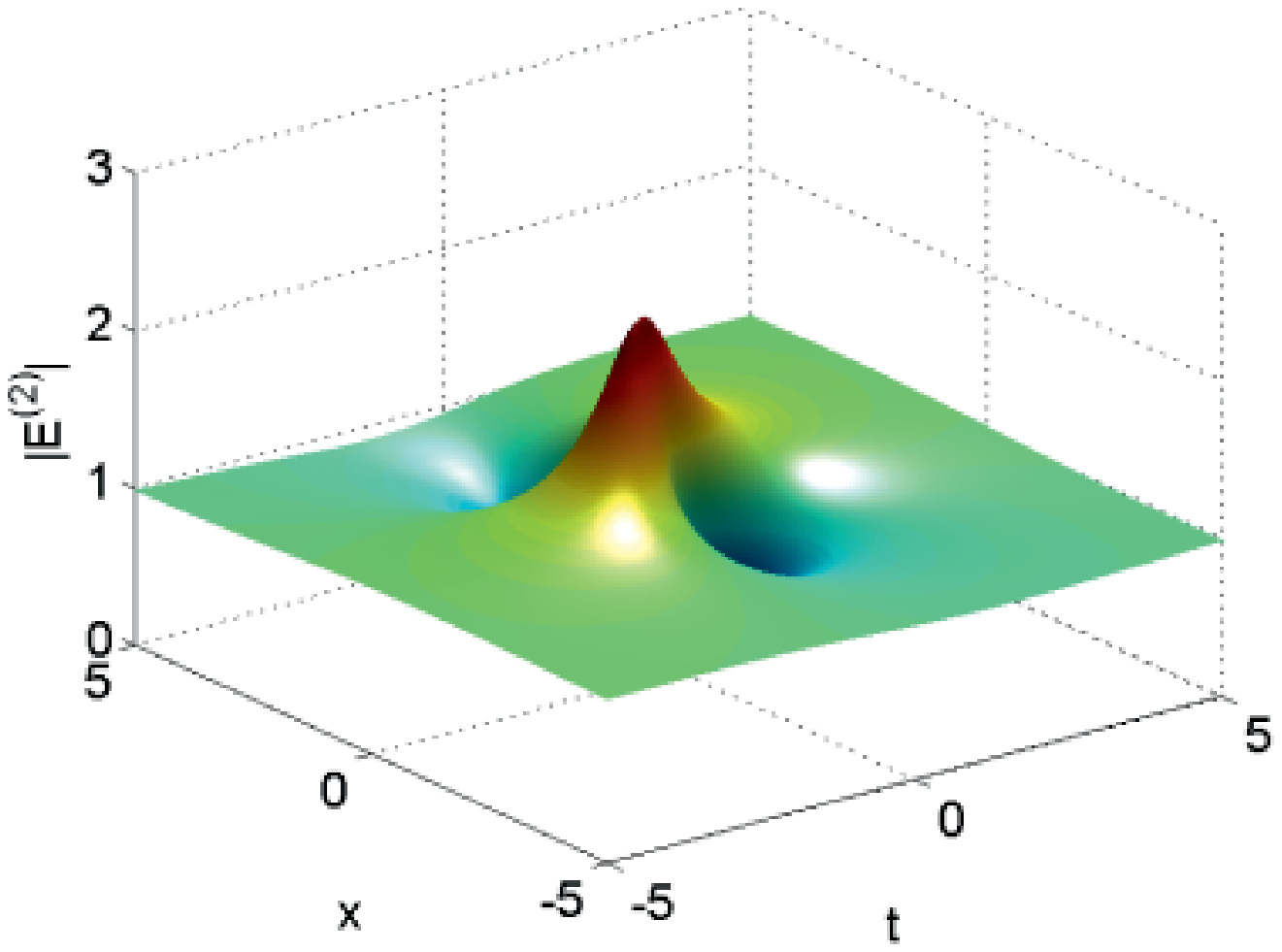}
    \end{center}
     \caption{Rogue waves envelope distributions $|E^{(1)}(x,t)|$
     and $|E^{(2)}(x,t)|$ of (\ref{pere}). Here, $a_1=3, a_2=1, q=1$.
		$k=2.36954 + 1.1972i$ and $\lambda=-1.69162 - 1.79721i$.
    } \label{fig_1}
\end{figure}

\begin{figure}[h]
\begin{center}
\includegraphics[width=7cm]{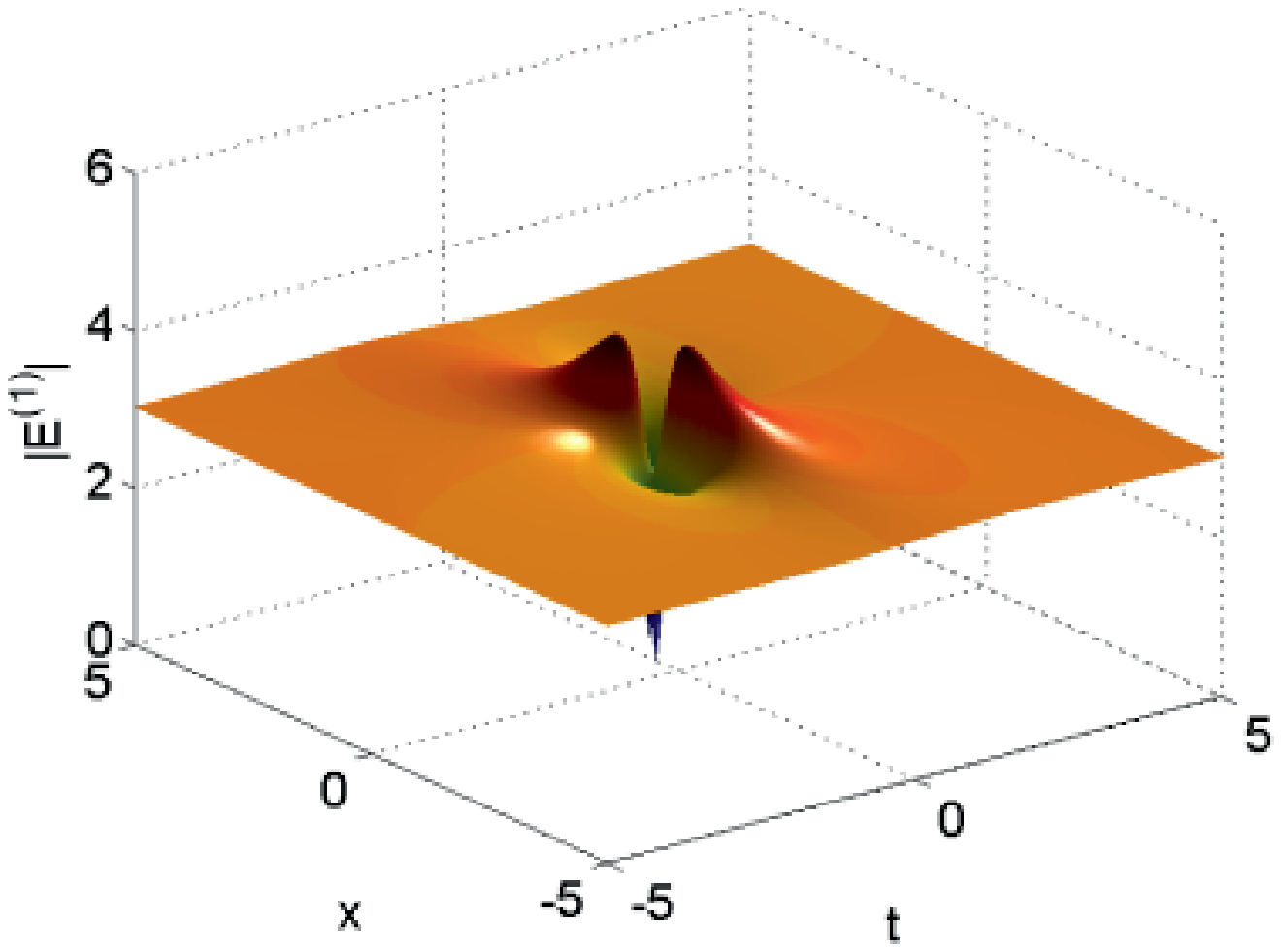}
\includegraphics[width=7cm]{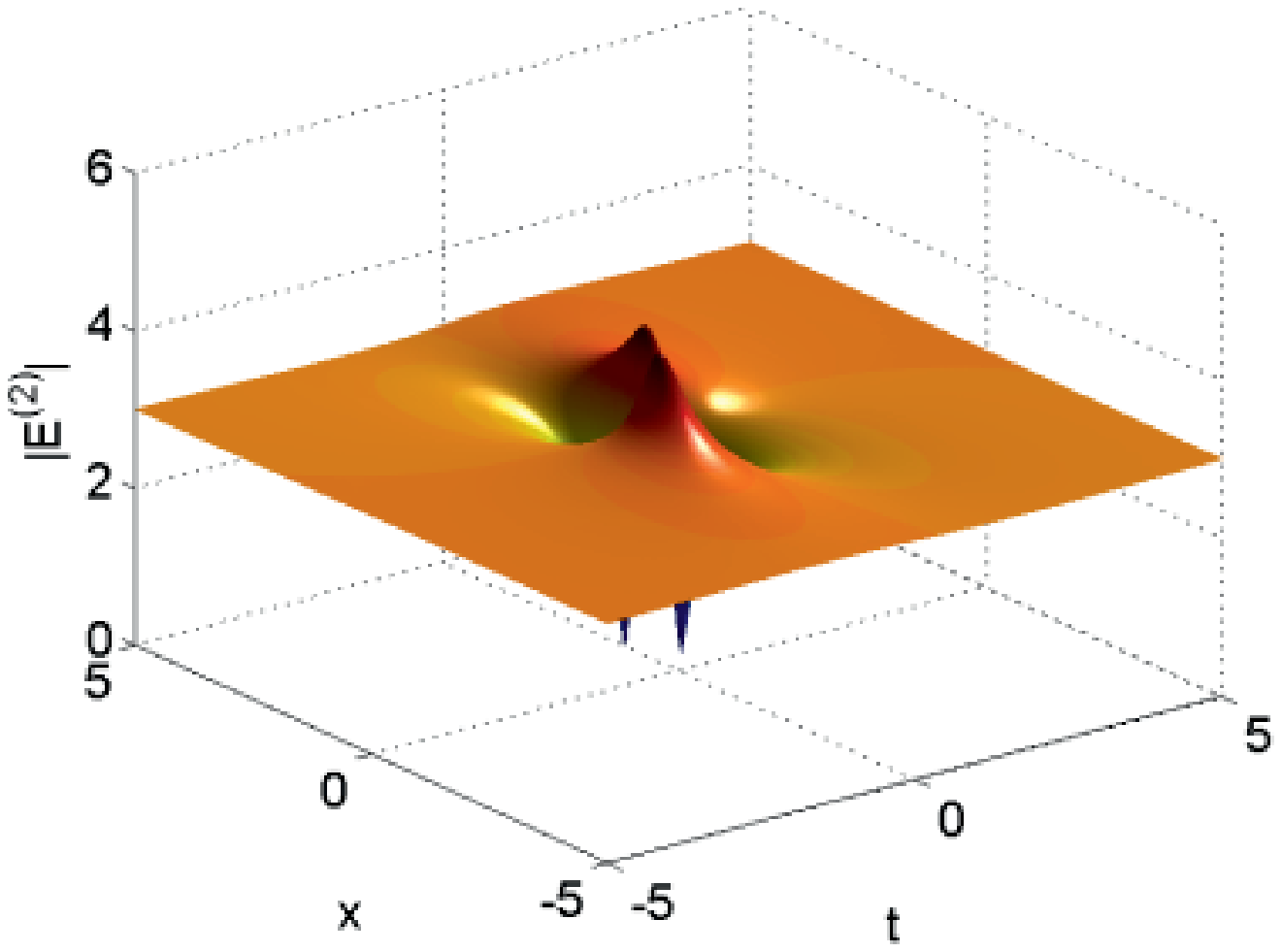}
    \end{center}
     \caption{Rogue waves envelope distributions $|E^{(1)}(x,t)|$
     and $|E^{(2)}(x,t)|$ of (\ref{pere}). Here, $a_1=3, a_2=3, q=1$.
		$k=4.02518i$ and $\lambda=-4.92887i$.
    } \label{fig_1bis}
\end{figure}

Figure \ref{fig_1} shows a typical dark-bright solution (\ref{pere}), while 
figure \ref{fig_1bis} shows a typical dark-dark solution (\ref{pere}). 

\begin{figure}[h]
\begin{center}
\includegraphics[width=7cm]{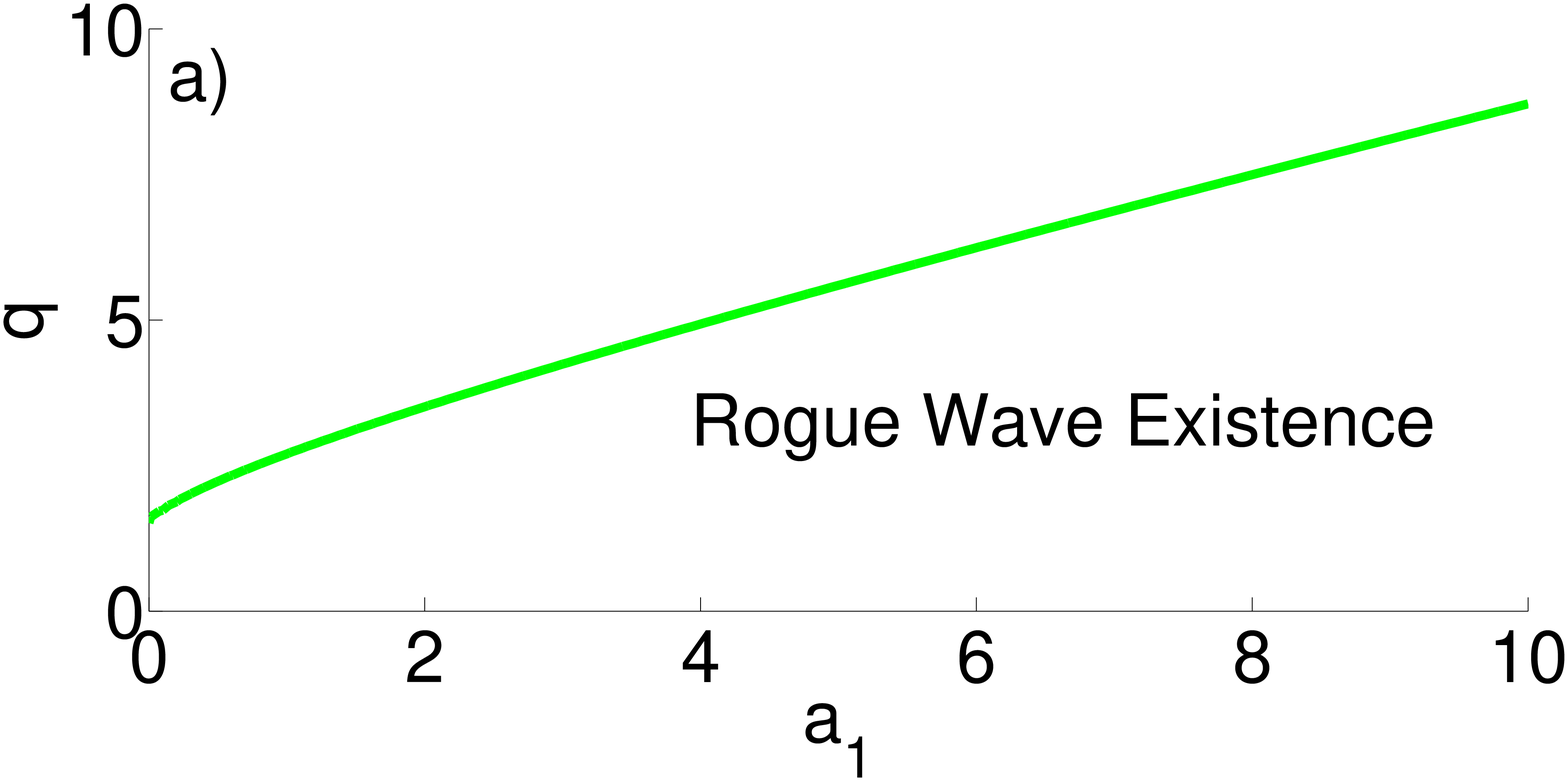}
\includegraphics[width=7cm]{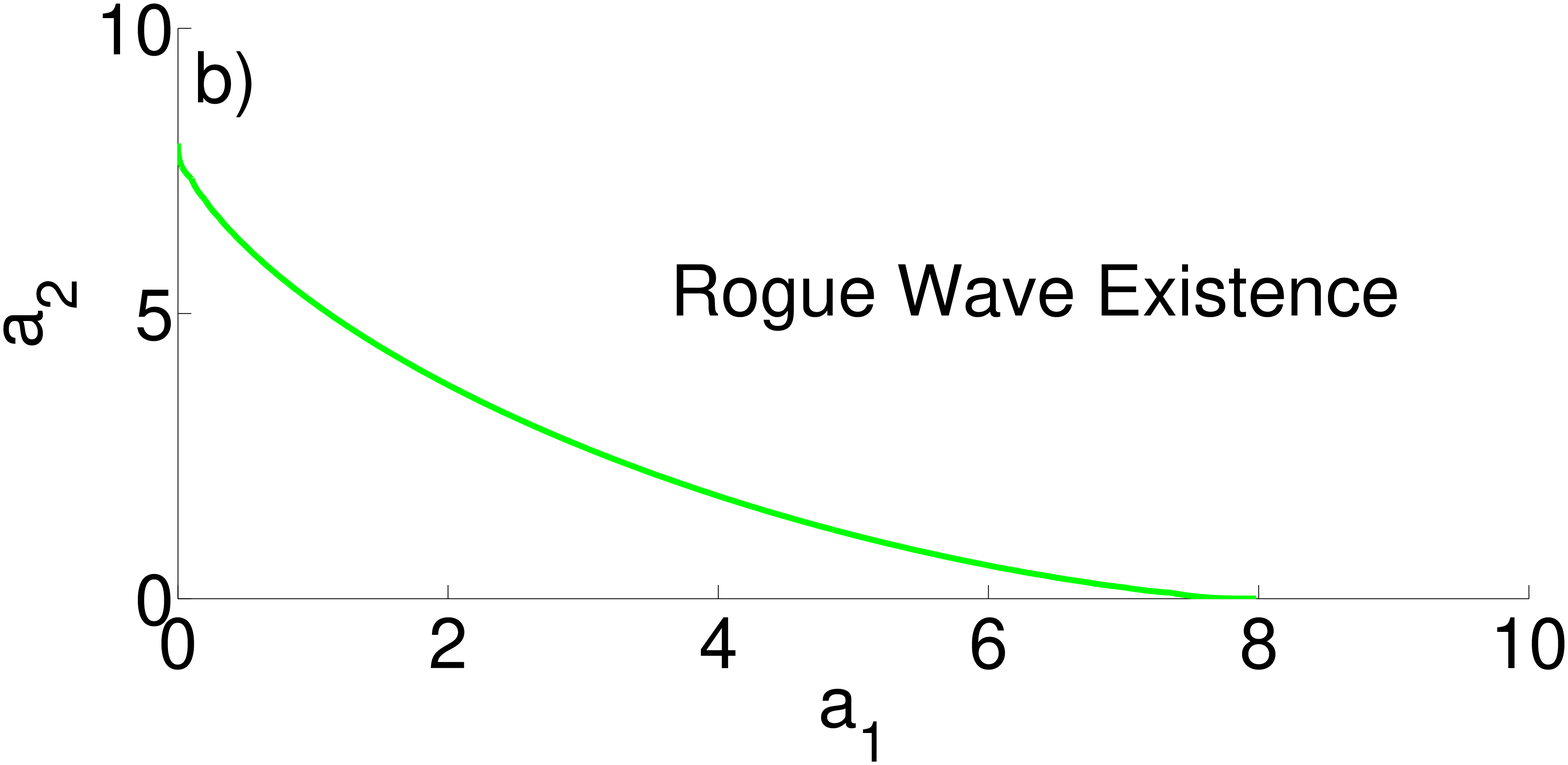}
    \end{center}
     \caption{Rogue wave existence condition {(\ref{eqkprima})}. a) $(q,a_1)$ plane,
		with $a_2=3$. b) $(a_2,a_1)$ plane, with $q=4$. } \label{figc}
\end{figure}

The family of solutions (\ref{pere}), found in the defocusing regime, possesses a novel 
feature with respect to families of solutions of Eqs. (\ref{VNLS}) previously reported in focusing regimes (see, f.i., 
\cite{baronio12,liu2013,zhai2013}). In fact, in the defocusing regime, threshold conditions 
for the parameters $a_1, a_2, q$ exist, due to the requirement that the solution $k$ of Eq. (\ref{eqk}) must be strictly complex, and that $\lambda$ is a double solution of Eq. (\ref{eql}).
We have identified these threshold conditions by computing the discriminant of
Eq. (\ref{eqk}). If this discriminant  is positive, Eq. (\ref{eqk})
possesses four real $k$ roots, and rogue waves do not exist; while if the discriminant 
is negative, Eq. (\ref{eqk}) has two real roots, to which no rogue wave is associated, and two complex conjugate $k$ roots, which instead imply the existence of rogue waves. This constraint on the sign of the discriminant leads to the following rogue wave existence condition:
\begin{equation} \label{eqkprima}
(a_1^2+a_2^2)^3 -12 (a_1^4 -7 a_1^2 a_2^2 +a_2^4) q^2 + 48 (a_1^2 + a_2^2) q^4 - 64 q^6 >0.
\end{equation}
Figs.  \ref{figc} a), b) report two characteristic examples of rogue waves existence conditions. In particular, Fig. 3 b) shows that, for fixed $q$, the background amplitudes have to be sufficiently large to allow for rogue wave formation. 
%
%
%
%In fact, the conditions on the parameters set 
%for the existence of a complex value of $k$, in Eq. (\ref{eqk}), and of a double solution of $\lambda$, 
%in Eq. (\ref{eql}), lead to thresholds conditions for the paraemters $q, a_1, a_2$.
%
{As a particularly simple example, consider the case $a_1=a_2=a$ and $q\neq 0$. In this case the inequality (\ref{eqkprima}) reads $4(a^2+4q^2)^2(2a^2-q^2) >0$, with the implication that 
%If $a^2 \leq q^2/2$, then the four zeros $k$, in Eq. (\ref{eqk}), are real and no complex values exist. If $a^2 > q^2/2$, then two zeros $k$ are real and the other two are imaginary with opposite sign:
%
%\begin{align*}
%	k=\pm i \Big(a^4-10a^2 q^2 -2 q^4+ |a| (a^2+4q^2)^{3/2}    \Big)^{1/2} ,
%\end{align*} 
%
%
%\begin{align}
	%\lambda=\frac{1}{12} [ \mp 4i + {(2-2i\sqrt{3}) (4+6a^2-3q^2)}/{[\pm (-8i-18i a^2 -18i q^2)+\sqrt{(4+6a^2-3q^2)^3-(-8i-18i a^2 -18i q^2)^2}]^{1/3}} + \\ 
	%-2i (-i+\sqrt{3}) [\pm (-8i-18i a^2 -18i q^2)+\sqrt{(4+6a^2-3q^2)^3-(-8i-18i a^2 -18i q^2)^2}]^{1/3} ],
%\end{align}  
%
%
%and the values of $\lambda$ result
%\begin{align}
	%\lambda=\frac{1}{12} \Big [ \mp 4i + \frac{(2-2i\sqrt{3}) l_0}{[\pm l_1+\sqrt{l_0^3-l_1^2}]^{1/3}} + \\ 
	%-2i (-i+\sqrt{3}) [\pm l_1+\sqrt{l_0^3-l_1^2}]^{1/3} \Big]  ,
%\end{align}  
%with
%%
%\begin{gather*}\label{def}
%%
%%
%%
%l_0 = 4+6a^2-3q^2,  
%l_1 = -8i-18i a^2 -18i q^2.
%\end{gather*}
%%
%\begin{align*}\lambda&=\frac{1}{12} [ \mp 4i + {(2-2i\sqrt{3}) l_0}{[\pm l_1+\sqrt{l_0^3-l_1^2}]^{1/3}} + \\ &-2i (-i+\sqrt{3}) [\pm l_1+\sqrt{l_0^3-l_1^2}]^{1/3} ], 
%\end{align*}  
%with $l_0 = 4+6a^2-3q^2,l_1 = -8i-18i a^2 -18i q^2$.
%
%\begin{gather*}\label{def}
%%
%%
%%
%l_0 = 4+6a^2-3q^2,  
%l_1 = -8i-18i a^2 -18i q^2.
%\end{gather*}
%
%
%
%Therefore, 
only in the (large amplitudes) subset $a^2 > q^2/2$ of the parameter plane $(a,q)$  
the rogue waves (\ref{pere}) do exist.}

%
%dashed white lines represent the approximate condition $2 |a_1 a_2| > q^2$. 

\textit{Defocusing VNLSE and induced MI.--}\label{sec3}
{Here we first turn our attention to the standard linear stability analysis of the background solution (\ref{fondo}), and then we prove
 that the existence of rogue waves 
is strictly related with a specific form of MI, namely to baseband MI.
In the following, the coupling parameter $s$ in (\ref{VNLS}) is considered as a continuous variable rather than a discrete, two-valued variable (i.e. $s=\pm 1$). Therefore, the background solution takes the expression 
$E_0^{(j)} =a_j e^{i(q_j x- \nu_j t)}$, with $\nu_j=q_j^2+2s(a_1^2+a_2^2),\; j=1,2$, while $a_1, a_2$ represent ``amplitudes'' (which, with no loss of generality, we take real valued), and $q_1, q_2$ represent ``frequencies''.
A perturbed nonlinear background can be written as 
$E_p^{(j)}=  [a_j+p_j] e^{i q_j x-i \nu_j t}$, where $p_j(x,t)$ are small perturbations (in amplitude and phase) which satisfy a linear equation.  Whenever $p_j(x,t)$ are $x-$periodic with frequency $Q$, i.e.,
$p_j(x,t)=\eta_{j,s}(t)e^{i Q x}+\eta_{j,a}(t)e^{-i Q x}$, their equations reduce to the $4\times4$ linear differential equation $\eta'=iM\eta$, with $\eta=[\eta_{1,s},\eta^*_{1,a},\eta_{2,s},\eta^*_{2,a}]^T$ (here a prime stands for differentiation with respect to $t$). For any given real frequency $Q$, the generic perturbation $\eta(t)$ is a linear combination of exponentials $\exp(i w_j t)$ where $w_j,\;j=1,\cdots,4, $ are the four eigenvalues of the matrix $M$. Since the entries $M_{mn}$ of the matrix $M$ are all real, $M_{11}=-Q^2-2 Q q_1 -2sa_1^2 $, 
$M_{22}=Q^2-2Qq_1+2sa_1^2$, $M_{33}=-Q^2-2Qq_2-2sa_2^2$, $M_{44}=Q^2-2Qq_2+2sa_2^2$,
$M_{12}=-M_{21}=-2sa_1^2$,
$M_{13}=M_{14}=M_{31}=M_{32}=-M_{41}=-M_{23}=-M_{24}
=-M_{42}= -2sa_1 a_2 $, $M_{43}=-M_{34}=2sa_2^2$, the eigenvalues $w_j$ are either real or come as complex conjugate pairs. They are the roots of the characteristic polynomial of the matrix $M$
\begin{equation} \label{eqP}
B(w)=w^4+B_3 w^3+ B_2 w^2+ B_1 w +B_0\;\;,
\end{equation}
with
\begin{gather*}\label{defk}
B_0 = (Q^2 - 4 q^2  ) [4 (sa_1^2 +  sa_2^2 - q^2) + Q^2] Q^4 \;;  \ \ \ \ \ \ \ \ \ \ \ \ \  \ \ \ \ \ \ \ \ \ \ \\
B_1 =  16 q (sa_1^2 - sa_2^2)  Q^3 \;; \ \ \ \ \ \ \ \ \ \ \ \ \ \ \ \ \ \ \ \ \ \ \ \  \ \ \ \ \ \ \ \ \ \ \ \ \ \ \ \ \ \ \ \ \ \ \  \\
B_2 = - 2  [2 (sa_1^2 + sa_2^2 + 2 q^2) + Q^2] Q^2\;; \ \ \ \ \ \ \ \ \ \ \  \ \ \ \ \ \ \ \ \ \ \ \ \ \ \ \ \ \ \ \ \ \ \ \\
B_3 = 0 \;.  \ \ \ \ \ \ \ \ \ \ \ \ \  \ \ \ \ \ \ \ \ \ \ \ \ \  \ \ \ \ \ \  \ \ \ \ \ \   \ \ \ \ \ \ \ \ \ \ \ \ \  \ \ \ \ \ \ \ \ \ \ \ \ \ 
\end{gather*} 
Whenever $M$ has an eigenvalue $w$ with negative imaginary part, $\textrm{Im}\{w\} < 0$, MI exists (see \cite{rot90,drum90,seve96,fato13} for details and review papers). Indeed, if the explosive rate is $G(Q) = -\textrm{Im}\{w\} >0$, perturbations  
grow exponentially like $\exp(G t)$ at the expense of the pump waves. 
The bandwidth of MI $0 \leq Q_1 < Q < Q_2$ in which $G(Q)\neq 0$ is baseband if $Q_1=0$ while is passband if $Q_1>0$. 
%The corresponding gain $G_m$ is defined as the maximum value taken by $G(Q)$ in the MI bandwidth.} 
%the standard linearized problem for the perturbation with the matrixelements $M_{m,n}$, 
%
%\begin{equation}
%M=\nonumber \left[\begin{array}{cccc}
%M_{1,1} & M_{1,2} & M_{1,3} & M_{1,4} \\
%M_{2,1} & M_{2,2} & M_{2,3} & M_{2,4} \\
%M_{3,1} & M_{3,2} & M_{3,3} & M_{3,4} \\
%M_{4,1} & M_{4,2} & M_{4,3} & M_{4,4} 
%\end{array}\right]
%\end{equation}
%
%$M_{1,1}=-Q^2-2 Q q_1 -2a_1^2 $, 
%$M_{2,2}=Q^2-2Qq_1+2a_1^2$, $M_{3,3}=-Q^2-2Qq_2-2a_2^2$, $M_{4,4}%=Q^2-2Qq_2+2a_2^2$,
%$M_{1,2}=-M_{2,1}=-2a_1^2$,
%$M_{1,3}=M_{1,4}=M_{3,1}=M_{3,2}=-M_{4,1}=-M_{2,3}=-M_{2,4}
%=-M_{4,2}= -2a_1 a_2 $, $M_{4,3}=-M_{3,4}=2a_2^2$.
%
%-Q^2-2 Q q_1 -2a_1^2 & -2a_1^2 & -2a_1 a_2 & -2a_1 a_2\\
%+2a_1^2 &  Q^2-2Qq_1 +2a_1^2 & +2a_1 a_2  & +2a_1 a_2 \\
%-2a_1 a_2 & -2a_1 a_2 & -Q^2-2Qq_2 -2a_2^2 & -2a_2^2\\
%+2a_1 a_2 & +2a_1 a_2 & +2a_2^2 & Q^2-2Qq_2 +2a_2^2
%
%
%Whenever $M$ has an eigenvalue with negative imaginary part, MI exists
%(see Ref. \cite{rot90,drum90,seve96,fato13} for details and review papers).

MI is well depicted by displaying the  gain $G(Q)$ as a function  of  $s$, $a_1, a_2$, $q_1$, $q_2$ and $Q$. Characteristic
outcomes of the MI analysis are reported in  Fig. \ref{fig_2}. 
\begin{figure}[h]
\begin{center}
\includegraphics[width=8cm,height=3cm]{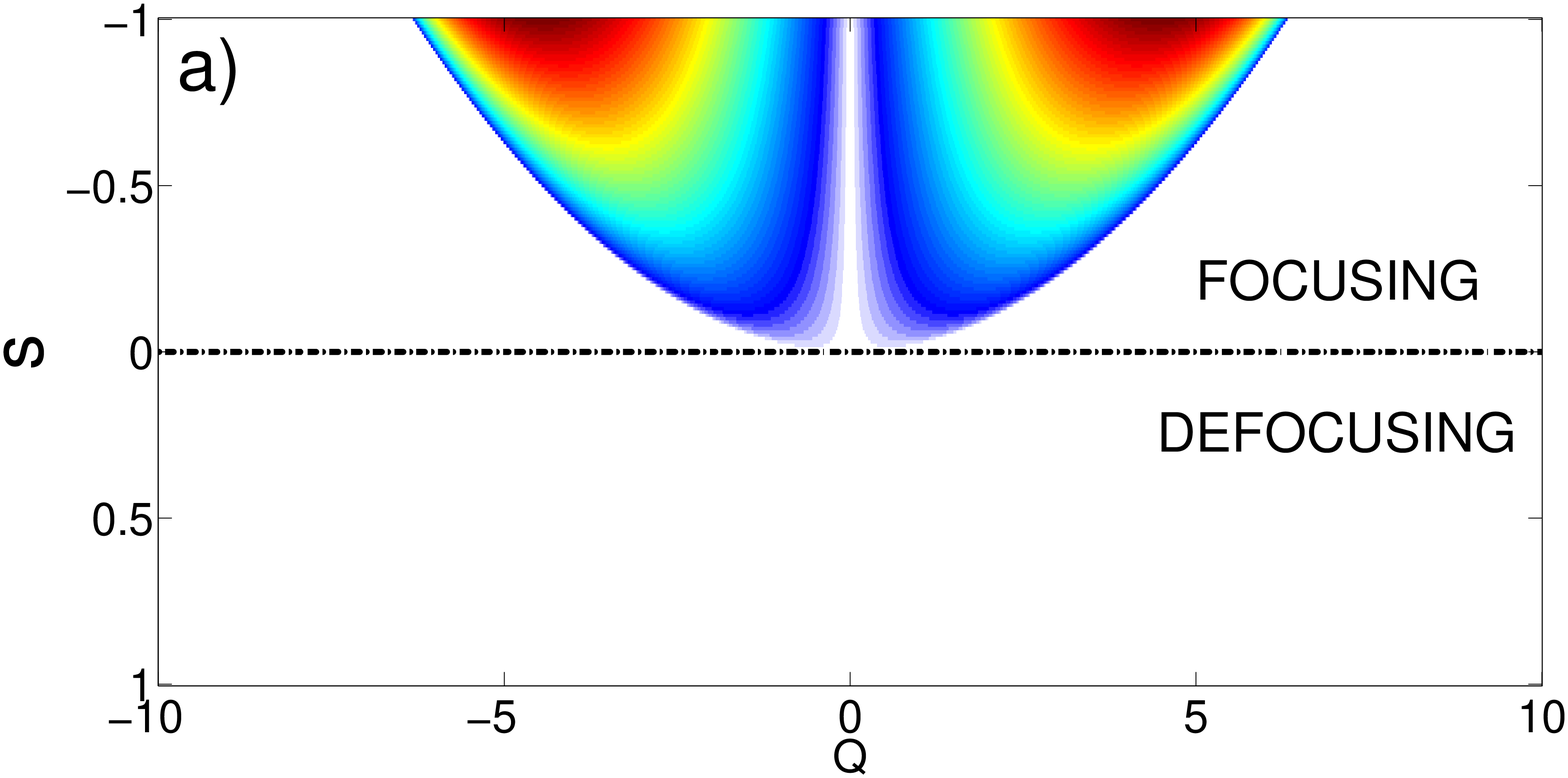}
\includegraphics[width=8cm,height=3cm]{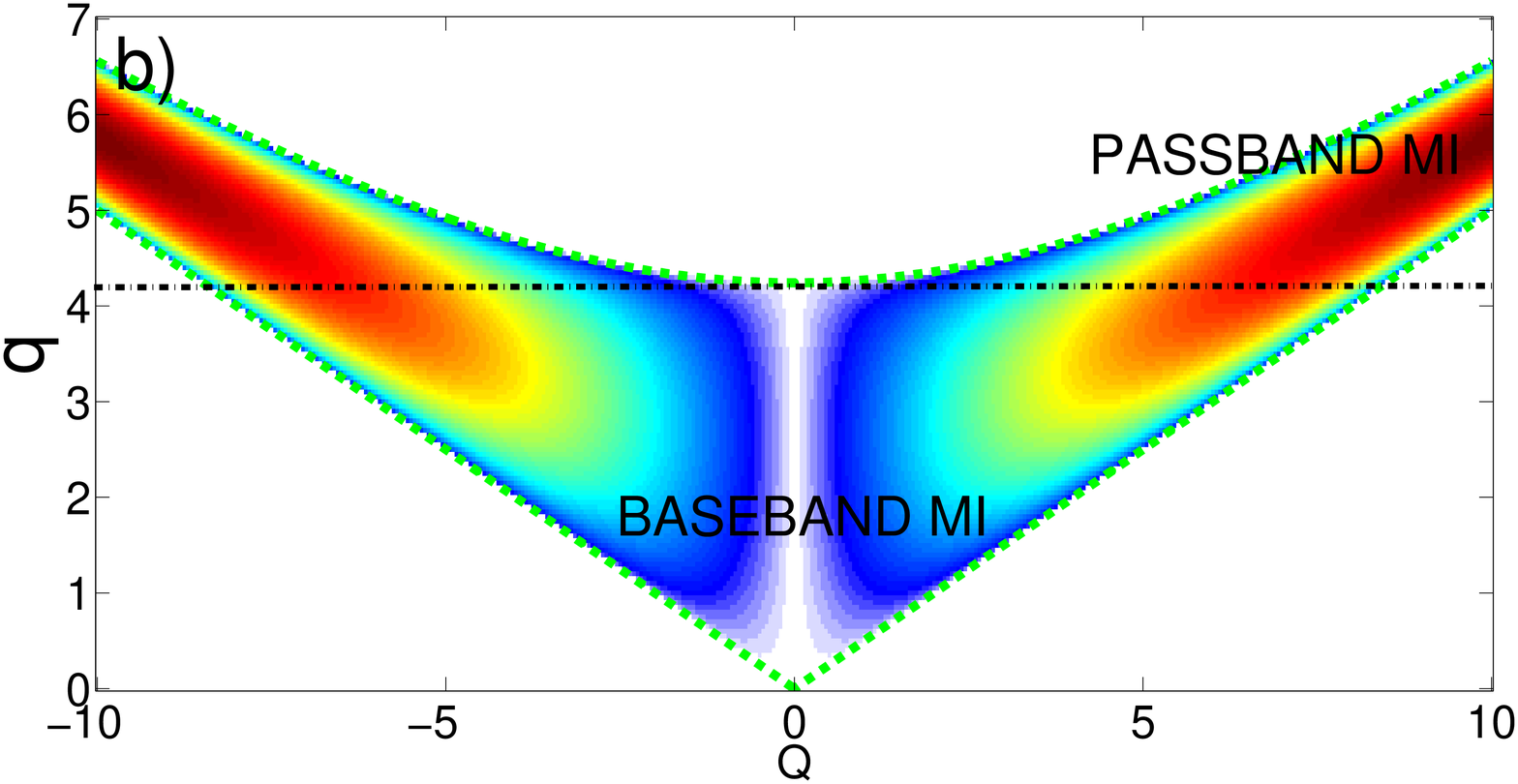}
\includegraphics[width=8cm,height=3cm]{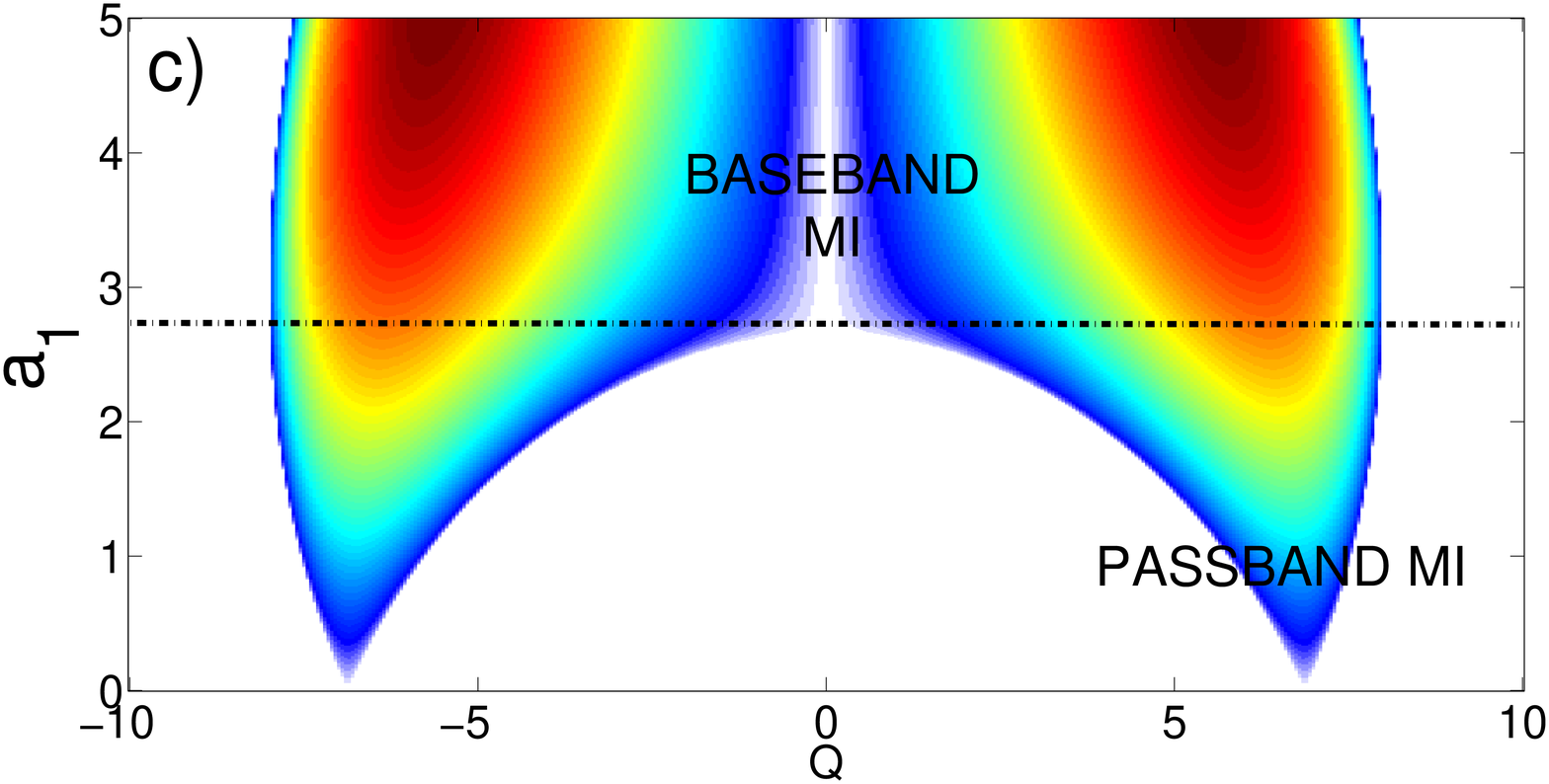}
    \end{center}
     \caption{Maps of MI of the VNLSE (\ref{VNLS}). a) MI on $(Q,s)$ plane,
		calculated for the case $a_1=3, a_2=1, q_1=q_2=1$. b) MI on the $(Q,q)$
		plane, calculated for the case $a_1=3, a_2=3$, $q_1=-q_2=q$ and $s=1$.  Dotted (green online) curves
		represent the analytical marginal stability condition $Q=2q$,
{$Q^2=\textrm{max}\{4 q^2 -8 a^2, 0\}$.}
		c) MI on the $(Q,a_1)$ plane, calculated for the case $a_2=3, q_1=-q_2=4$
		and $s=1$.
    } \label{fig_2}
\end{figure}

Figure  \ref{fig_2} a) corresponds to the case where the nonlinear background modes have the same
frequencies ($q_1=q_2$, thus $q=0$). In this case, MI is always present in the focusing regime
($s<0$), but it is absent in defocusing regime ($s>0$). This particular case corresponds 
to the trivial vector generalization of scalar NLSE MI dynamics. We remark that no
rogue waves exist in this defocusing regime.

Figure \ref{fig_2} b) corresponds to the case where the nonlinear background modes have opposite frequencies
($q_1=-q_2=q$), in a defocusing regime $s=1$ which yields MI. 
The higher $q$, the higher $G$. In the special case of equal background amplitudes $a_1=a_2=a$, 
the marginal stability conditions can be found analytically: {$Q^2=4q^2$, 
$Q^2=\textrm{max}\{4 q^2 -8 a^2, 0\}$. Thus, for $a^2 > q^2/2$ a baseband or lowpass MI, which
includes frequencies that are arbitrarily close to zero, is present (i.e. $0 < Q^2 < 4q^2$). Instead, 
for $a^2 \leq q^2/2$, the MI occurs for frequencies in the passband range $(4q^2-8a^2) < Q^2 < 4q^2 $.
We remark that in the previous section we have shown that rogue waves (\ref{pere}) necessarily exist for $a^2 > q^2/2$
(f.i. the parameters of the rogue wave of Figs. (\ref{fig_1bis}) correspond to the baseband 
MI as shown in Fig. \ref{fig_2} b)). Thus, rogue waves (\ref{pere}) and baseband MI coexist.
Figure \ref{fig_2} c) corresponds to the case in which the  nonlinear background modes have different
frequencies ($q_1=-q_2=q$), and different input amplitudes $a_1\neq a_2$ in the defocusing regime $s=1$.
For low values of $a_1$, only passband MI is present. By increasing $a_1$, the baseband MI condition is eventually reached.
\newline 
Thus, we proceed by focusing our interest on the MI behavior in the limit $Q \rightarrow 0$, namely on the occurrence of baseband MI. To this aim, we rewrite the characteristic polynomial (\ref{eqP}) as $B(Qv)=Q^4b(v)$ and consider the polynomial $b(v)$ at $Q=0$, namely 
\begin{equation} \label{eqp}
 b(v)=v^4+b_3 v^3+ b_2 v^2+ b_1 v +b_0,
\end{equation}
with
\begin{gather*}\label{defc}
b_0 = -16 q^2 (a_1^2 +  a_2^2 - q^2)\;;\;  b_1 =  16q (a_1^2 - a_2^2) \;;   \\
b_2 = - 4 (a_1^2 + a_2^2 + 2 q^2) \;; \;  b_3 = 0.  
\end{gather*}
Next, we have evaluated the discriminant of  (\ref{eqp}). If the discriminant is positive, the polynomial (\ref{eqp}) possesses four real roots, 
and no MI occurs; while if the discriminant  is negative, Eq. (\ref{eqp}) possesses 
two real roots and two complex conjugate roots, and Eqs.(\ref{VNLS}) exhibit baseband MI.
\newline
The interesting finding is that the previous sign constraint on the discriminant of the polynomial (\ref{eqp}), which leads to the baseband MI condition, turns out to coincide with the sign constraint (\ref{eqkprima}) which is required for rogue wave existence.}
{These results are important as they show that i) the rogue wave solutions (\ref{pere}) exist
in defocusing regimes in the subset of the parameters space where MI is present; ii) the rogue waves solutions (\ref{pere}) exist if and only if baseband MI is present.}

%
%scrivere meglio figure, commentare, curve di stabilità argiael e commenti su passband o baseband
%   

%
%collegamento con rogue waves
%
%
\textit{Conclusions.--}\label{sec4}
{We presented and analyzed exact, explicit rogue-wave solutions of the defocusing VNLSE.
This family of solutions includes both bright and dark  components. Moreover, we
clarified that the rogue wave existence condition is strictly related to a very specific manifestation of MI, namely MI whose bandwidth includes arbitrarily small frequencies. %bound to induced MI conditions, in particular to baseband MI conditions. 
The existence of rogue wave solutions in the
defocusing regime is expected to be crucial in explaining
extreme waves in a variety of practical multi-component defocusing systems, 
from oceanography to optics and plasma physics.}

%\textit{Acknowledgement. }
The present research was supported by the Italian Ministry of University and 
Research (MIUR, Project Nb.2009P3K72Z, Project Nb. 2012BFNWZ2), by the  Italian 
Institute for Nuclear Physics (INFN Project Nb. RM41), by the Agence Nationale
de la Recherche (ANR TOPWAVE), by the Netherlands Organisation for Scientific 
Research (NWO, Grant 639.031.622), and by ONR (Grant Nb. 214 N000141010991).
%$^*$Corresponding author. fabio.baronio@unibs.it

%\clearpage


\begin{thebibliography}{}
\bibitem{spiegazione}
The computation of rogue wave solutions is based on the
Darboux technique applied to the Lax pair associated
with the VNLSE. This method is well known and does
not need to be detailed here to any extent. The relevant
literature is rather vast and we refer to \cite{deg09}
for the formalism and general set up adopted here, and to
Ref. \cite{baronio12,dega13} (and
to the literature quoted therein) for the basic arguments
to follow for the construction of rational solutions.
\end{thebibliography}
\end{document}